\documentstyle[manuscript,aps]{revtex}
\draft
\begin{document}
\title{Ballistic resistivity in aluminum nanocontacts}
\author{A. Hasmy$^1$, A.J. P\'{e}rez-Jim\'{e}nez$^2$,
J.J. Palacios$^2$, P. Garc\'\i a--Mochales$^{3}$,\\
J.L. Costa--Kr\"{a}mer$^{4}$, M. D\'\i az$^{1}$, E. Medina$^1$,
and P.A. Serena$^{3}$}
\address{$^1$\it Centro de F\'\i sica, IVIC, Apdo. 21827,
Caracas 1020A, Venezuela}
\address{$^2$\it Departamento de F\'\i sica Aplicada,
Universidad de Alicante,\\
San Vicente del Raspeig, 03690--Alicante, Spain}
\address{$^3$\it Instituto de Ciencia de Materiales de
Madrid,
CSIC,
Cantoblanco, 28049--Madrid, Spain}
\address{$^4$\it Instituto de Microelectr\'{o}nica de Madrid,
CSIC,
Isaac Newton 8, PTM,\\ 28760--Tres Cantos, Madrid, Spain}
\maketitle
\vskip-1cm
\begin{abstract}
One of the major industrial challenges is to profit from some fascinating
physical features present at the nanoscale. The production of dissipationless
nanoswitches (or nanocontacts) is one of such attractive applications.
Nevertheless, the lack of knowledge of the real efficiency of electronic
ballistic/non dissipative transport limits future innovations.
For multi-valent metallic nanosystems -where several transport channels per
atom are involved- the only experimental technique available for statistical
transport characterization is the conductance histogram. Unfortunately its
interpretation is difficult because transport and mechanical properties are
intrinsically interlaced. We perform a representative series of
semiclassical molecular dynamics simulations of aluminum nanocontact
breakages, coupled to full quantum conductance calculations, and put in
evidence a linear relationship between the conductance and the
contact minimum cross--section for the geometrically favored aluminum
nanocontact configurations. Valid in a broad range of conductance values,
such relation allows the definition of a transport parameter for nanomaterials,
that represents the novel concept of ballistic resistivity.
\end{abstract}
\newpage

If the mean free path of electrons is larger than the nanocontact
size, the transport supported by propagating channels is expected
to be ballistic (i.e., for aluminum, at room temperature, the
electron mean free path is 45 nm, whereas at 4K it is of the order
of milimiter).
For contact sizes of the order of a few Fermi
wavelengths $\lambda_F$, well defined modes (channels) appear
associated to the transversal confinement of electrons. In such
limit, the conductance $G$ is well described by the Landauer
formula $G=G_0\sum_{n=1}^NT_n$, where $G_0$=2e$^2$/h is the
conductance quantum (e being the electron charge and h Planck's
constant), $T_n$ is the transmission probability of the $n$-th
channel, and $N$ is the number of propagating modes with energies
below the Fermi energy\cite{Landauer1970}. Previous
studies\cite{Scheer1998} demonstrated that, in principle, the
number of conducting channels are determined by the number of
valence electrons of the respective chemical element, and the
transport efficiency or transmission probability of electrons can
differ significantly depending on the nanocontact
structure\cite{Jelinek2003}. While for monovalent noble--metals
such as Cu, Ag and Au, the transmission probability $T$ has been
estimated to be approximately equal to 1 (i.e. at a nanocontact
neck, each noble metal atom contributes with $G_0$ to the
conductance value\cite{Rodrigues2000,Lee2004}), for some
monovalent alkali--metals or polyvalent chemical species, the
channel transmitivity can result smaller than one in single--atom
contacts\cite{Scheer1998,Lee2004,Cuevas1998}.
The few
detailed studies available in the literature relate to
single--atom contacts\cite{Scheer1998}. It has been evidenced that
the only electronic channel available to a single--atom gold
contact has a better transmission performance than any single
electronic channel, of the possible three, for the aluminum
atom\cite{Scheer1998,Rodrigues2000}. Thus, more channels
do not guarantee better transport at the atomic scale.
The transmission
efficiency has its origin in the scattering processes taking place
associated with the particular electronic structure of the atoms
forming the nanocontact region. Then, existing ballistic theories
that relate the conductance with the contact size (namely, the
Sharvin formula\cite{Sharvin1965} with its semiclassical
corrections within the free electron model\cite{Torres1994}), fail
because they neglect the 'chemistry' in nanoconstrictions.

Electronic transport measurements on metallic nanocontacts of
different sizes, have been made possible by means of scanning
tunnelling microscopy (STM)\cite{Olesen1994}, mechanically
controllable break--junctions (MCBJ)\cite{Agrait1993,Krans1995},
or simply separating two macroscopic wires in table--top
experiments\cite{Costa95}.
By indenting one electrode into another and then separating them,
one observes a stepwise decrease in the electrical conductance, until the
breakpoint is reached,
corresponding to the formation of a single--atom nanocontact.
Each scan of the conductance dependence on elongation
differ from one another,
since structural evolutions during breakages are not identical\cite{Hasmy2001}.
However, statistically, the
accumulation of data from many scans gives rise to a histogram of
peaked structures, a clear evidence for the
existence of preferential conductance values.
Such conductance histograms are a robust reproducible characteristic for a
given metal species under fixed experimental parameters (such as temperature
and applied voltage). On the other hand, recent molecular dynamics
simulations of a series of aluminum nanocontact breakages,
reveal a peaked structure corresponding to preferred geometrical
configurations at the nanocontact neck\cite{Hasmy2001}. In spite
of the quantum features of conductance at the atomic scale, these
results suggest that a relation exists between preferred atomic
configurations and the conductance histogram peaks\cite{Medina2003,Yanson1999,Yanson2001}.

We have implemented a state--of--the art Embedded Atom
Molecular Dynamics method for the simulations of aluminum
nanocontact rupture\cite{Hasmy2001,Medina2003,Mishin1999}. The atoms
were initially distributed in a supercell formed by 18 layers
perpendicular to the (111) fcc direction, containing 56 atoms
each. The lattice constant is initially taken to be 4.05 $\AA$.
The direction (111) corresponds to that in which the
contact is elongated until breakdown. Simulations are performed at
4 K. In a first stage the system is relaxed during 50 picoseconds.
After this relaxation, two bilayer slabs are defined at the top
and bottom of the relaxed supercell, and are separated at a
velocity of 2 m/s. The atoms inside these slabs are frozen
during subsequent stages, defining the bulk supports of the
nanocontact during the breaking process. The other atoms move and
reaccommodate into new configurations during the
elongation process. The full determination of atomic positions
during contact stretching allows the evaluation of the
evolution of its minimum cross--section $S_m$. The determination
of $S_m$ has been done by considering a standard numerical
procedure, which is able to determine the nanocontact slice with
the smallest cross--section $S_m$ in number of atoms
\cite{Bratkovsky1995}. The slice thickness is assumed to be equal
to the covalent atom diameter. With this methodology, $S_m$ can result
in a non integer value when, for instance, the contact atomic layers
are irregular.

Following a similar strategy to that of nanocontact transport
experiments, we performed many numerical realizations of wire
breakages for the statistical analysis of the conductance. For all
scans, and resulting configuration each 20 ps, we computed the
conductance using a full quantum mechanical procedure based on the
{\it ab initio} Gaussian embedded--cluster
method\cite{Palacios2002}. Due to computer time limitations, the
conductance was computed for configurations with $S_m \le 5$
restricting the quantum calculation to a nanocontact region formed
by 5 atomic layers, describing the narrower (and most important,
in terms of electronic transport) nanocontact section. This narrow
layer is formed by the minimum cross section layer (where $S_m$ is
evaluated) and its two neighboring layers below and above. A
similar strategy has been recently proposed\cite{Dreher2004} for
constructing computational gold conductance histograms, although a
parameterized Tight-Binding approach has been used to calculate
conductance values.

Typical evolution of the minimum cross--section $S_m$,  and the
corresponding conductance $G$, during the nanocontact breakage are
shown in Figure 1. The shapes of the curves reveal the existence
of a strong correlation between conductance and the nanocontact
neck section size, in agreement with previous numerical results
for other
materials\cite{Rodrigues2000,Bratkovsky1995,Nakamura1999}. A
striking fact is that for values $S_m\approx 1$ (defining a
nanocontact of one--atom section) there are two different
conductance values ($G/G_0 \approx 2$ and $\approx 1$). These
values correspond to different atomic
arrangements\cite{Jelinek2003}. On the one hand, the monomer
contact configuration (see Figure 1a) provides conductance values
of the order of $G/G_0 \approx 2$, while on the other hand, the
dimer contact configuration (see Figure 1b) gives rise to
conductances close to $G/G_0 \approx 1$. This conductance
bi--valuation, for the $S_m \approx 1$ case, shows that the
orbital valence accommodates differently depending on the contact
coordination and on the separation between the contact atom and
its neighbors, a finding that confirms previous
observations\cite{Jelinek2003,Lang1995}. Also, we have noticed
that when a neck section of two--atom contact lifts under
stretching, a dimer--chain contact is formed, giving rise to a
conductance jump from a value greater than 3 to 1.2 (see Fig. 1b).
With increasing stretching (time in the figure), $G$ slowly
decreases and then increases from 0.90$G_0$ to 1.05$G_0$. This
occurs when the dimer--chain contact evolves from a position
perpendicular to the (111) direction towards a parallel alignment
with this direction (which is perpendicular to the supporting
slabs). Such increase of $G$ before the rupture, reproduces STM
transport measurements\cite{Cuevas1998b}, confirming the validity
of our model, and reveals the possible improvement of electronic
resonant conditions with strain, as predicted by previous
numerical calculations\cite{Lee2004,Cuevas1998b}. Finally, when
the contact breaks, one reaches the tunnelling regime and the
conductance falls to 0 as the nanocontact ceases to stretch.

For more than 800 aluminum nanocontact configurations (obtained
from the evolution of 50 stretching sequences), we estimated the
conductance per minimum cross--section $G_a=G/S_m$, which is
equivalent to the sum of the three channel transmitivities
available for aluminum atoms at the narrowest neck section. Figure
2 depicts $G_a$ as a function of the minimum cross--section $S_m$.
In spite of the data dispersion, the figure suggests that $G_a$
converges to a constant value as $S_m$ increases. We recall that
the independence of $G_a$ on the contact size has been previously
observed experimentally for gold nanocontacts\cite{Rodrigues2000},
where it was observed that $G_a=G_0$, which is the maximum
possible value for the atomic conductance of any monovalent
material \cite{Rodrigues2000,Yanson2001,Nakamura1999}. For the
aluminum case, Fig. 2 shows that for almost all configurations,
$G_a$ results larger than $G_0$, and converges to a value between
$G_0$ and 1.5$G_0$. Then, the data suggests that aluminum
nanocontacts are, per atom, better conductors than any monovalent
metal wire.

As usually done in
experiments\cite{Medina2003,Yanson1999,Yanson1997,Diaz2001,Halbritter2002},
we accumulated all $G$ traces and constructed the first
computational aluminum conductance histogram (see Fig. 3, top).
The good agreement between this numerical result and previously
published experiments, again evidences the good fidelity of our
calculations, i.e. aluminum conductance histogram peaks are close
to integer multiples of the conductance
quantum\cite{Yanson1997,Diaz2001,Halbritter2002}. We recall that a
peaked structure has been also observed in minimum cross-section
histograms\cite{Hasmy2001,Dreher2004}, reflecting the existence of
energetically favorable atomic configurations at the nanocontact
neck. Regarding experiments, the main advantage, when performing
molecular dynamics simulations, is that we can separate the
corresponding conductance contributions in nanocontact
configuration families, grouping configurations that posses a
similar number of atoms at the narrowest nanocontact
cross-section. In order to identify such configuration families,
we use the integer label $N_a$ ($N_a$ = 1,2,3...) to describe the
set of configurations with minimum cross-section $S_m$ comprised
between $N_a - 1/2$ and $N_a + 1/2$. This label is equivalent to
the effective number of atoms defining the narrowest nanocontact
region. Partial conductance histograms concocted in the previous
fashion are depicted in Fig. 3 and show that for all $N_a$ values,
there corresponds a peak at a conductance value $G_{M}(N_a)$. For
comparison, we also show in Figure 3 the average conductance
$\langle G (N_a) \rangle$ for each partial conductance histogram
(see vertical dashed lines). Note that for the monoatomic contact
case ($N_a = 1$), two conductance peaks appear at $G/G_0=1$ and
$G/G_0=2$, which correspond, respectively, to the conductance
contributions of the dimer--chain (see inset of Fig. 1b), and the
single--central one--atom (see inset of Fig. 1a) contact
configurations. For $N_a \ge 2$ the average conductance $\langle G
(N_a) \rangle$ is close to the peak position $G_{M}(N_a)$
indicating that conductance distributions are rather symmetrical
around its maximum value.

Figure 4 plots the quantity $G_M (N_a)/N_a$ (open circles)
as a function of the effective number of atom contacts $N_a$.
In this Figure, we have included two points corresponding to particular
aluminum nanocontact configurations
with $N_a = 20$ and 30.
The convergence of the atomic conductance to a constant value between
1 and 1.5 is now much more evident than in Figure 2.
Figure 4 also include a plot of the
conductance maximum $G_{M}(N_a)$ as a function of $N_a$
(gray squares, corresponding label figure appears at the right hand side).
The good quality of a linear fit of the data ($\chi^2$=0.9995)
suggests that the conductance converges to a straight line
(with a slope of 1.16 $G_0$).
Additionally, we plot the average conductance $\langle G (N_a) \rangle$ for each
contact configuration family $N_a$ (small black circles in Fig. 4),
and its corresponding standard error.
Within the estimated error bars, the $\langle G (N_a) \rangle$
values reflect the same linear behavior observed for
$G_M (N_a)$.
Therefore it is evident that a linear relation between the conductance
and the effective number of atom contacts $N_a$ can be established.
The slope of the curve is now
an electronic structure specific property that can be bundled into a peculiar {\it ballistic resistivity}
defined by
$$G=N_a/\rho_b$$
where $N_a$ is dimensionless and $1/\rho_b$ has the dimensions of conductance.
Here, $\rho_b$ retains the value of 0.86 $R_0$, where $R_0$ denotes the quantum
unit resistance, or 12907 $\Omega$.

Semiclassical approximations based on the free electron model have
a fundamental shortcoming in the face of the previous result: They
only depend on geometry, and the Fermi wavelength is the only
material dependent parameter, while backscattering due to both
geometrical and electronic structure constraints are the main
culprit for the appearance of ballistic resistance in
nanomaterials\cite{Scheer1998,Nakamura1999}. In our conductance
calculations, the scattering phenomenon is implicitly described in
the considered quantum methodology. In that sense the new
ballistic resistivity cannot be interpreted as {\it ohmic}, but a
result of a novel scaling behavior of the material conductance at
the mesoscale. It is this scaling that promises to be a universal
material independent property. In such transport regime, where the
mean free path of electrons is larger than the sample size, the
conductance dependence on the contact length loses meaning, while
the dependence on the cross--sectional size is preserved as in
many theoretical approximations (i.e the Sharvin conductance is
proportional to the contact minimum cross--sectional area $A$). In
our work, the notion of the area $A$, instead of $N_a$, is
inadequate since any consideration defining it in a complex
electronic structure of few atom contacts results speculative.
Notwithstanding it is intuitive that a first order approximation
for such an area should behave linearly with the effective number
of atom contacts $N_a$.

Finally, one should expect that the described ballistic
resistivity $\rho_b$ will also depend on thermodynamical
variables, as the the minimum
cross--section histograms depend on temperature\cite{Hasmy2001}. Such properties, and
the extension of this kind of studies on other chemical elements
with different electronic structures, seem to conform a field of a
promising research activity, due to the obvious attractive
applications of these knowledge in the emerging nanoelectronic
industry.

\vskip 1cm
{\bf Acknowledgments.} We thank J. J. S\'aenz for helpful
discussions, and Cecalcula (Venezuela) for computer facilites.
This work has been partially supported by the CSIC-IVIC
researchers exchange program and the Spanish DGICyT (MEC) through
Projects MAT2000-0033-P4 and BFM2003-01167/FISI.

Correspondence \ and \ requests \  for materials \ should \ be \ addressed \ to
\ A.H. (e-mail: anwar@ivic.ve).

\newpage

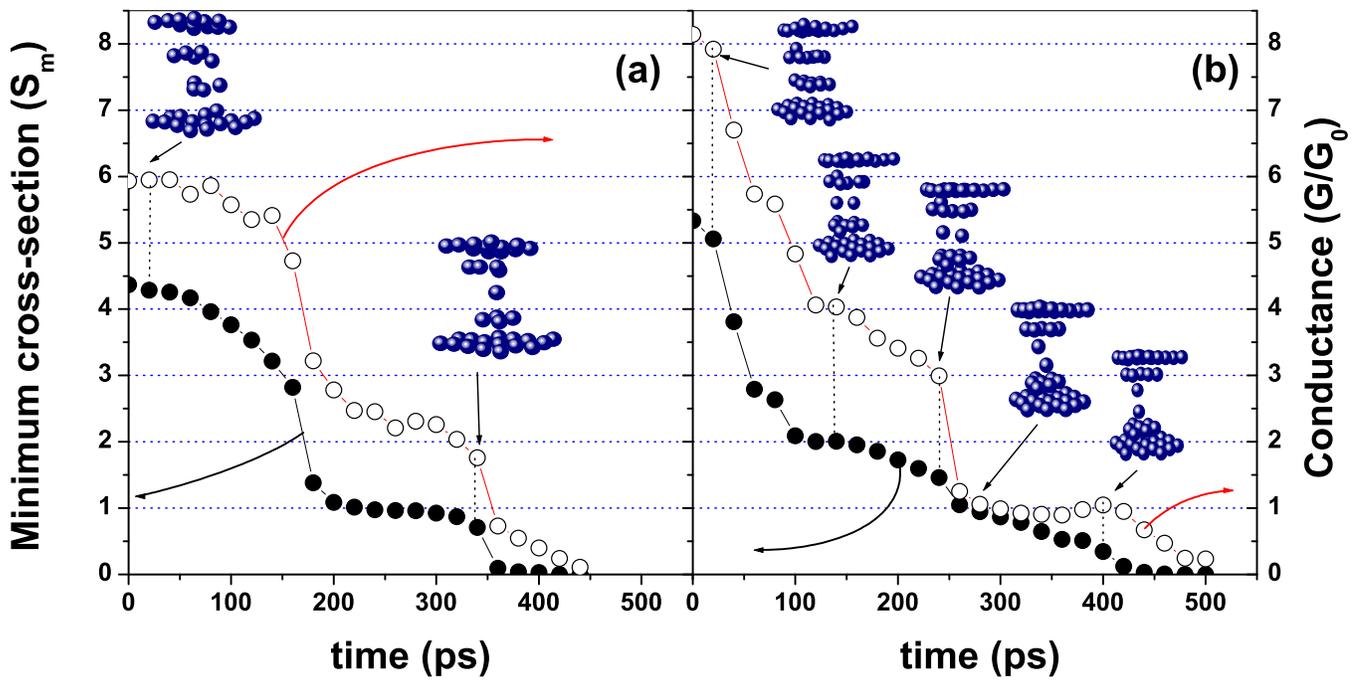
\begin{figure}[ht]
\caption{Figures (a) and (b) illustrate two examples of the time
evolution of the minimum cross--section (black symbols) and the
quantum conductance (open symbols) during aluminum wire breakages.
The results are obtained from a combination of Embedded Atom
Molecular Dynamics and ab-initio Gaussian Embedded--cluster
method, respectively. The temperature is equal to 4K. The insets show
some of the atomic configurations at the contact region. (a)
illustrates a rupture mechanism which passes through a single--central
atomic configuration before breakage. (b) illustrates the formation
of a dimer--chain contact before breakage. Arrows and dashed vertical lines
denote the corresponding conductance and minimum cross--section
associated to these nanoneck configurations. Note that the
dimer--chain and the single--central atom contact have a similar
minimum cross section value ($S_m \approx 1$), but the conductance
results equal to $G_0$ and 2$G_0$, respectively.
The stretching mechanism in (b) involves a rotation of a dimer unit
at the nanoneck.
For larger necks, the results suggest that there is a proportionality factor between
the minimum cross--section and the conductance.}
\label{fig1}
\end{figure}

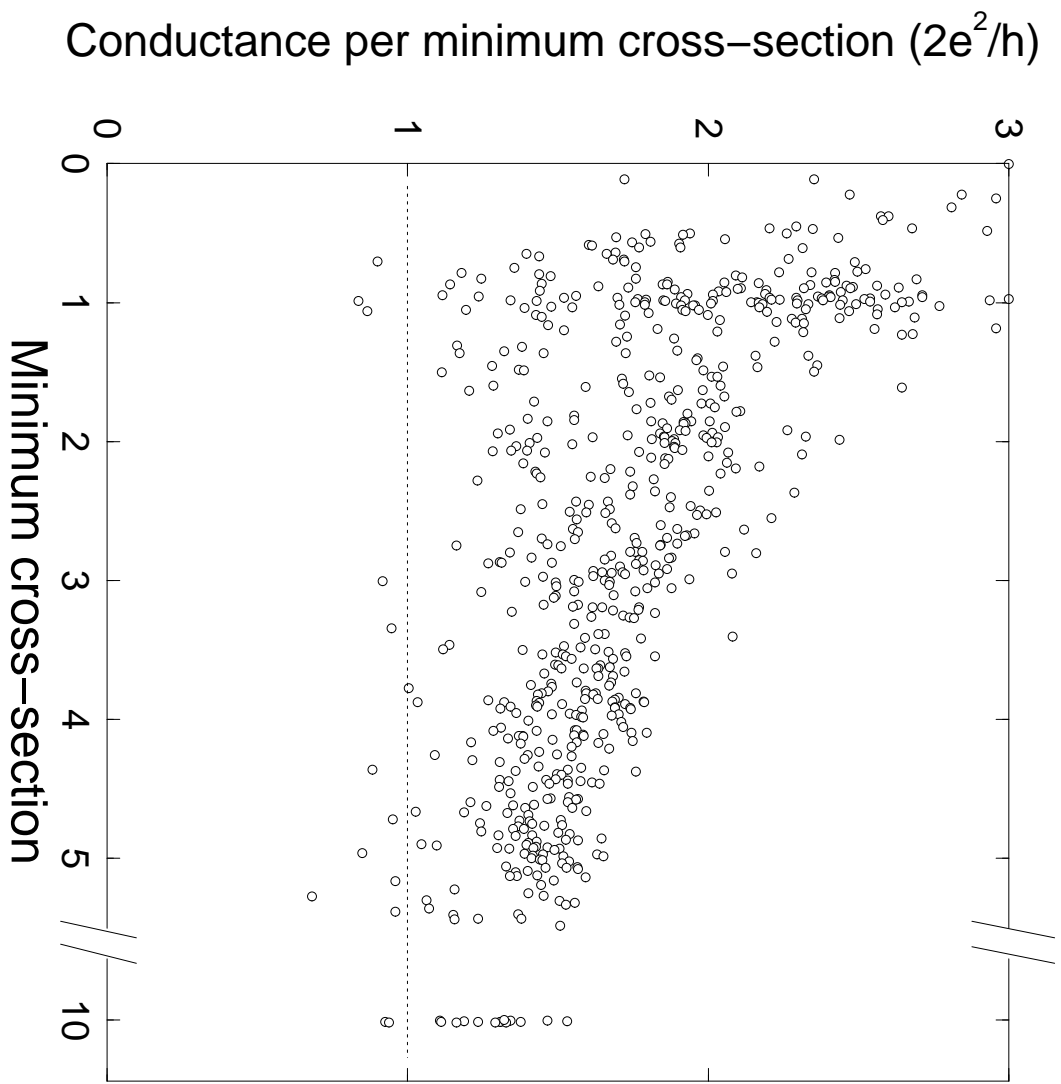
\begin{figure}[ht]
\caption{The conductance per minimum cross--section $G_a$
(i.e. the quantum conductance divided by the minimum cross--section)
plotted as a function of the minimum cross-section for all
configurations, results from fifty simulated aluminum nanocontact
breakages. In spite of the
data dispersion, it is observed that such atomic conductance is
larger than $G_0$ and converges to a constant value as the
contact size increases. The dotted line is a guide to the eye,
which denotes the quantum conductance value.} \label{fig2}
\end{figure}

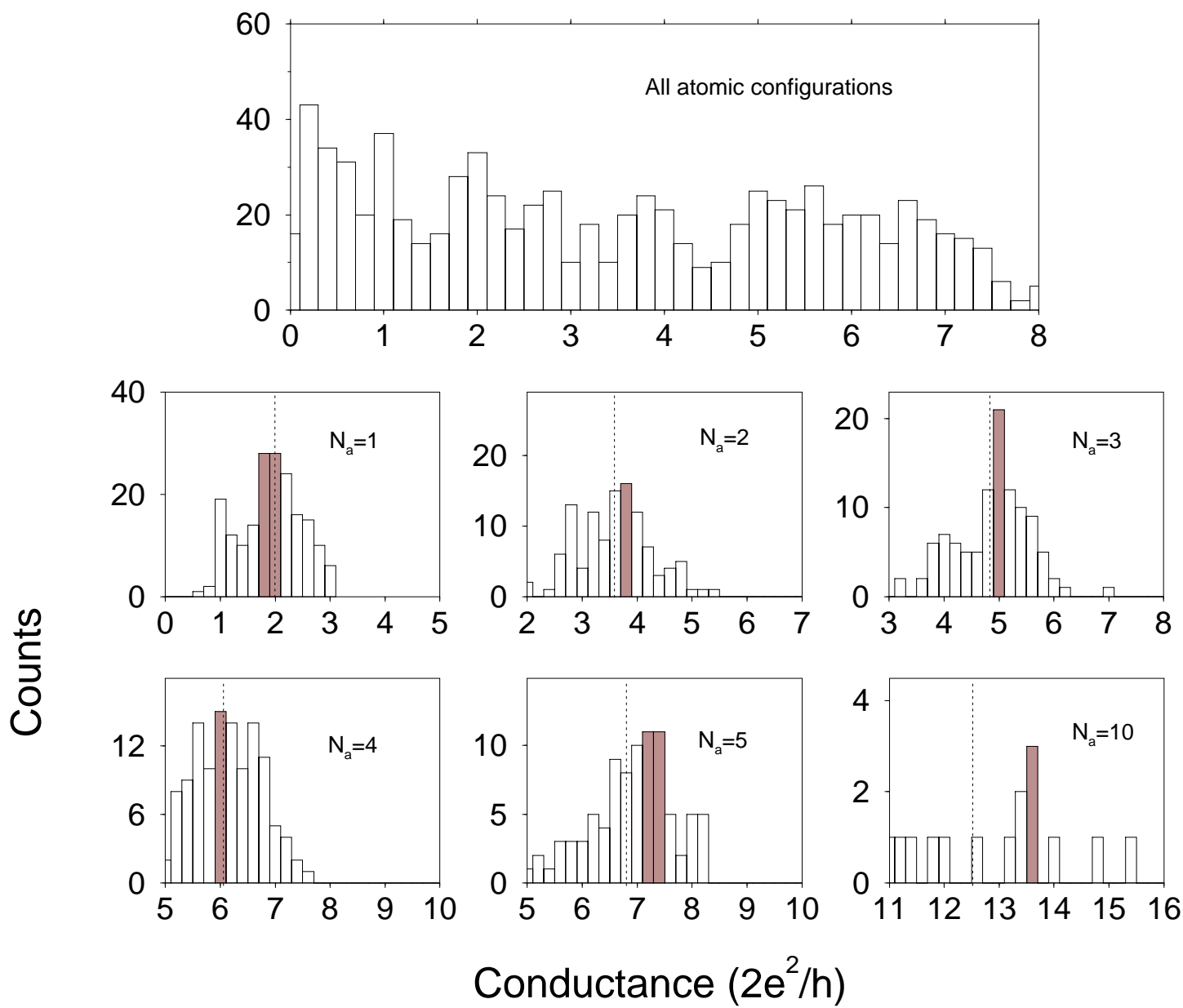
\begin{figure}[ht]
\caption{The top of the figure shows the calculated conductance
histogram for aluminum. It includes more than 800 nanocontact
configurations. The peak structure of the histogram is reminiscent of what
has been previously observed in aluminum conductance experiments.
Below this histogram, we separately show the contribution to the
global conductance histogram of configurations corresponding to effective number
of atoms $N_a$= 1, 2, 3, 4, 5 and 10. Such number of atoms is
equivalent to the calculated minimum cross--section value, but in
integer precision. We observe that in each histogram there appears
a maximum at a conductance value $G_M (N_a)$ (bins corresponding
to these maxima were filled with gray.
Vertical dashed lines show the computed average conductances $\langle G(N_a) \rangle$
for each depicted conductance distribution.} \label{fig3}
\end{figure}

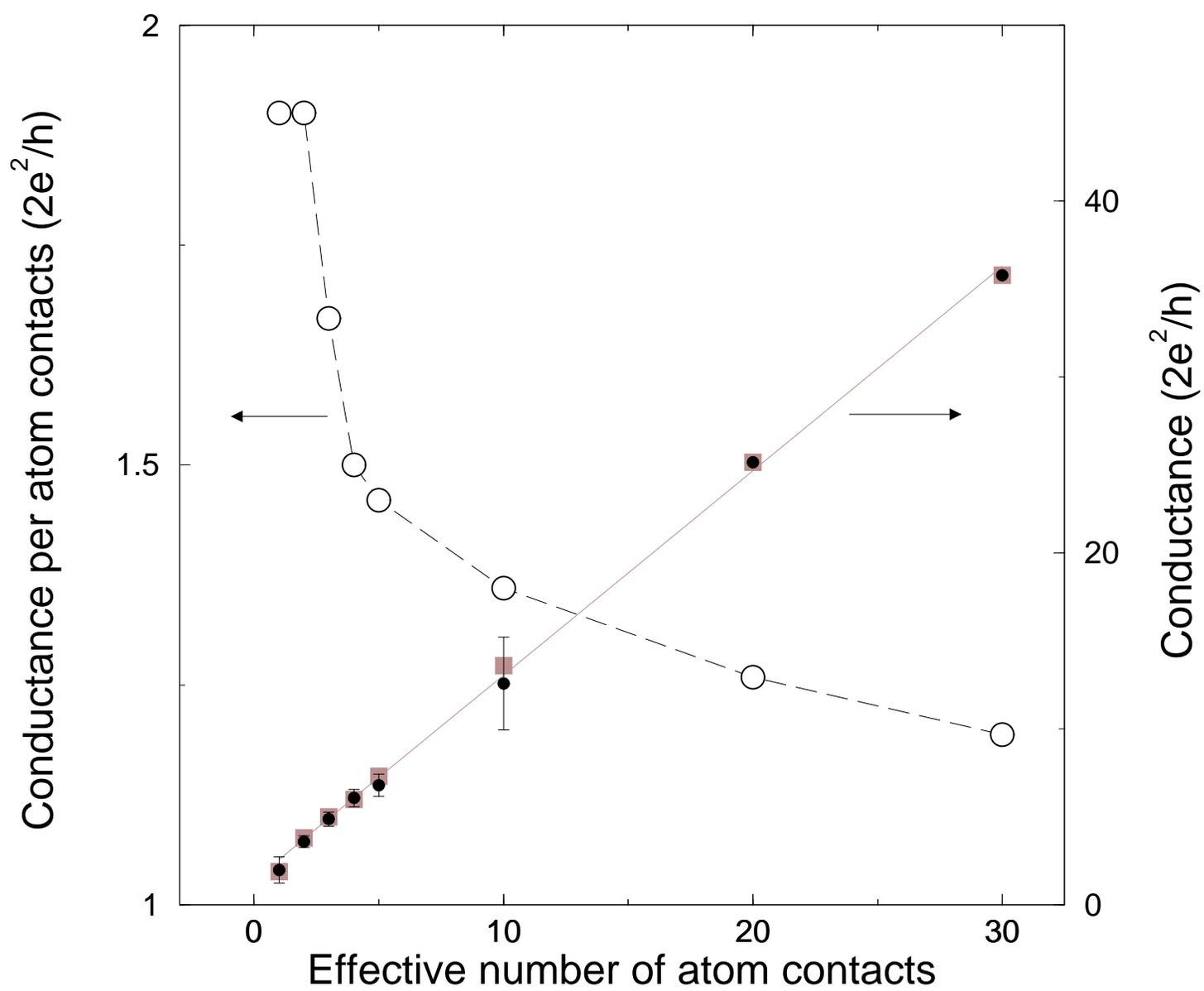
\begin{figure}[ht]
\caption{The conductance per atom is plotted as a function of the
effective number of atom contacts $N_a$ (open circles). Here, the convergence  to
a constant atomic conductance value is more evident than in Fig.
2. The figure label at the right hand side refers to the
conductance $G_M (N_a)$ curve (gray symbols) for the respective number of contact
atoms. A linear fit gives a slope equal to
1.16, suggesting that this is the value for which the atomic
conductance should converge for larger nanocontact sizes.
The figure also shows the corresponding average conductance
$\langle G(N_a) \rangle$ for each
value of the effective number of atom contacts (black circles, the bars
indicate the respective standard error of the conductance mean),
and shows that this estimation is also consistent with the linear fit.
The results demonstrate the proportionality between conductance
and number of atoms $N_a$, and reveals the existence of a transport
parameter which does not depend on the contact size. The inverse
value of the slope (0.86) denotes an effective ballistic
resistivity of preferable atomic configurations for aluminum
nanocontacts.} \label{fig4}
\end{figure}

\end{document}